\documentclass[intlimits,twoside,a4paper]{article}

\usepackage{amsmath,amssymb}
\usepackage{graphicx}

\usepackage[T2A]{fontenc}
\usepackage[cp1251]{inputenc}

\usepackage{graphics}  

\newcommand{\mean}[1]{\left\langle{#1}\right\rangle}

\usepackage[eqsecnum]{cmpj3}

\articletype{Rapid Communication}

\issue{2019}{22}{2}{24001}
\doinumber{10.5488/CMP.22.24001}

\title[Stochastic particulate matter and air pollutants dynamics]%
{Taking drift-diffusion analysis from the study of turbulent flows to the 
study of particulate matter smog and air pollutants dynamics}
\author[T. Varapongpisan, L. Ingsrisawang, T.D. Frank]{T. Varapongpisan\refaddr{label1}, 
L. Ingsrisawang\refaddr{label1}, 
T.D. Frank\refaddr{label2,label3}}
\addresses{
\addr{label1}
Department of Statistics, Faculty of Science, Kasetsart University,
Chatuchak, Bangkok, 10900, Thailand\\
\addr{label2} 
CESPA, Department of Psychology, 
University of 
Connecticut, 406 Babbidge Road, Storrs, CT 06269, USA\\
\addr{label3} 
Department of Physics,
University of 
Connecticut, 2152 Hillside Road, Storrs, CT 06269, USA
}

\date{Received March 14, 2019, in final form April 22, 2019}

\begin{document}

\maketitle

\begin{abstract}

Drift-diffusion analysis has been introduced in physics 
as a method to study turbulent
flows. 
In the current study, it is proposed to use the method to identify
underlying dynamical models of particulate matter smog, ozone and
nitrogen dioxide concentrations.  
Data from Chiangmai are
considered, which is a major city in the northern part of Thailand that
recently has witnessed a dramatic increase of hospitalization that are assumed
to be related to extreme air pollution levels. 
Three variants of the drift-diffusion analysis method
(kernel-density, binning, linear approximation) are considered. 
It is shown that all three
variants explain the annual pollutant peaks
during the first half of the year by assuming that the parameters of the 
physical-chemical evolution
equations of the pollutants vary periodically throughout the
year. 
Therefore, our analysis provides evidence
that the underlying dynamical models
of the three pollutants being considered are explicitly
time-dependent. 

\keywords drift-diffusion analysis, particulate matter, air pollutants 
\pacs 02.50.Ey, 05.10.Gg, 05.40.-a, 92.60.Sz 
\end{abstract}

An important task of nonlinear physics and statistics is 
to identify the underlying mechanisms that determine the evolution of systems
on the basis of experimental data.
In this regard, in physics, 
a method has been developed to investigate
turbulent flows~\cite{friedrich97prl}
that is nowadays frequently called drift-diffusion analysis. The method
was in part motivated by the self-similarity hypothesis of turbulent flows
that in its own merit has been investigated in various systems (see e.g.,
\cite{khomenko10fnl}).
In recent years,
various studies have 
examined turbulence
using the drift-diffusion analysis 
approach~\cite{friedrich97physd,naert97pre,tutkun04,kuwahara17},
see also \cite{friedrich11}.
However, the method turned out to have a broad spectrum of applications (for a
review see~\cite{friedrich11}). 
For example, sport and movement sciences have been taken advantage of
drift-diffusion analysis to identify movement- and posture-related 
dynamical
systems~\cite{davieskurz13,kurz13,mourik06physletta,frank06pre,gottpeinkelippensnagel09pla}. 
Bistable lasers~\cite{frank05jnpcs,chiangga10} and engineering
problems~\cite{gradisek00pre,lindpeinke14,noiray17} have been
examined. 
In what follows, the drift-diffusion analysis approach will be used
to identify underlying laws determining the evolution of air pollutants.
Those laws are assumed to reflect the relevant physical-chemical
evolution equations of the pollutants under consideration as well as
 the impacts of meteorological conditions. 
Monthly extreme value data of air pollutants will be considered
because such extreme air pollutant concentrations 
are assumed to come with serious health 
risks~\cite{lid18,maji17,pope06}
and are likely to increase death  
rates~\cite{giri07,monterolorenzo11}. We will analyze data from 
the city of Chiangmai,
Thailand. While Chiangmai is not the largest city of Thailand, it is the
largest city in the northern part of Thailand. Importantly, in recent years,
the number of hospitalizations that are due to high air pollutants 
concentrations is dramatically
increasing in Thailand, in general, and in Chiangmai, in 
particular~\cite{vichit11}. Therefore, a better understanding of the
dynamics of the monthly extreme scores of air pollutant concentrations would
be beneficial. We will consider the following three air pollutants:
particulate matter that is of 10 micrometers of less (PM$_{10}$), ozone
(O$_3$), and nitrogen dioxide (NO$_2$).

Our departure point is a time-discrete sequence of 
observations of pollutant concentrations. This sequence will be 
 referred to as historical trajectory $X^{h}(n)$ given for the
time points $n=1,\dots,N$ (with $N=60$, see below). 
In what follows, $n$ will denote
consecutive months. Our goal
is to derive a stochastic model from the historical trajectory in
analogy to the proposal by Friedrich-Peinke-Renner for historical financial
data and to take seasonal effects into account.
Following the Friedrich-Peinke-Renner method~\cite{friedrich00aprl}, we
consider the increments $Y_n(\tau)$ defined by
$Y_n(\tau) = X(n+\tau) - X(n) \Rightarrow X(n+\tau)=Y_n(\tau) + X(n)$
for $\tau\geqslant 0$ and $Y_n(0)=0$. Parameter $\tau$ defines a time scale.
The increments $Y_n$ are assumed to satisfy an evolution equation 
that describes
how
$Y_n(\tau)$ evolves
from small scales of a few months (e.g., $\tau \approx 1,2,3$) 
to large scales of a
year (e.g., $\tau \approx 12,13,14$).
In order to determine that evolution equation, 
we consider $R$ increment trajectories of length $S$ with
$\tau = 0,\dots,S$,  $n=1,\dots,R$, and $R+S=N$. The evolution equation for
$Y_n$ is then obtained using the drift-diffusion 
analysis~\cite{friedrich97prl}.

Although drift-diffusion 
analysis~\cite{friedrich97prl}
is as such a non-parametric data analysis method,
it requires to fix \textit{a priori} the type of the stochastic model
under consideration. In what follows,
we consider a model given in terms of the  
stochastic iterative map 
\begin{equation}
\label{eq2}
Y_n(\tau+1)= f\big(Y_n(\tau),m(n,\tau)\big) + g\big(m(n,\tau)\big) \epsilon(\tau) \ .
\end{equation}
In equation~(\ref{eq2}),
$f$ will be referred to as drift function
(in analogy to the drift function of
a Fokker-Planck equation~\cite{risken89book,frank05book}). 
The drift function $f$ is assumed to depend on 
the month $m$ of the year,
where
$m$  depends on $n$ and $\tau$
like
$m(n,\tau) = v  \ \mbox{if}  \ v \in [1,11]$
and 
$m(n,\tau) = 12  \ \mbox{if}  \ v=0$
with $v=(n+\tau) \ \mbox{mod} \ 12$.
In equation~(\ref{eq2}) 
$\epsilon(\tau)$ denotes statistically 
independent random variables distributed like a 
normal distribution with mean zero and variance $2$. 
The parameter
$g\geqslant 0$ is the noise strength or noise amplitude and, in general, may
depend on the month of the year.
Moreover, $g^2$ is the noise variance. For the sake of simplicity, it is assumed
that $g$ does not depend on the state $Y_n$ (i.e., an additive noise model
is considered). In early studies by
Friedrich and Peinke~\cite{friedrich97prl}
and Stanton~\cite{stanton97}
on the drift-diffusion
analysis, Friedrich, Peinke, and Stanton  
have determined 
representations for drift and diffusion coefficients
of Markov diffusion processes
in terms of conditional averages. In analogy to those
representations, from equation~(\ref{eq2}) we obtain the
Friedrich-Peinke-Stanton representation of the drift
in terms of the conditional average
\begin{equation}
\label{eq4}
f(z,m) = \mean{Y_n(\tau+1)}\big|_{Y_n(\tau)=z} .
\end{equation}
For the noise variance we obtain
\begin{equation}
\label{eq5}
g^2(m)=\frac{1}{2}\mean{ \big[Y_n(\tau+1)-f\big(Y_n(\tau),m\big)\big]^2}  , 
\end{equation}
which is not a conditional average because we assume that the noise term 
is
state-independent (i.e., 
additive).
The drift function $f$ can 
approximately be described by means of several methods.
The Friedrich-Peinke binning 
method~\cite{friedrich97prl,friedrich97physd}
yields the estimator 
\begin{equation}
\label{eq6}
f(z,m) \approx \sum_{j=1}^K c_j(m) \chi_j(z) , \qquad 
c_j(m) = 
\frac{1}{Z_{jm}} 
\sum_{\tau=0}^S \left[
\sum_{n=1,Y_n(\tau) \in I_j}^R 
\delta_{mn} Y_n(\tau+1)
\right]  , 
\end{equation}
where $\chi_j$ are indicator functions equal to 1 in appropriately defined
intervals.
We consider $Y_n \in [A,B)$ and use $K$ bins of width $\Delta y$
such that $y_1=A$, $y_{K+1}=B$, and $y_j=A+(j-1)\Delta y$. The bin-intervals
are $I_j=[y_j,y_{j+1})$. The indicator function is $\chi_j(z)=1$ if 
$z \in I_j$ and $\chi_j(z)=0$ otherwise.
In equation~(\ref{eq6}), $\delta_{mn}$ is the Kronecker function
that equals 1 if the (running) month $n$ corresponds to a particular month $m$
of the year and zero otherwise. That is, only those
pairs $Y_n(\tau)$, $Y_n(\tau+1)$ contribute to $c_j(m)$ 
for which the (running) month $n$ is the month of the year $m$
of interest. Moreover, we have
$Z_{jm}=
\sum_{\tau=0}^S [
\sum_{n=1,Y_n(\tau) \in I_j}^R 
\delta_{mn}]$.
The kernel density estimation method suggested by 
Stanton~\cite{stanton97}
yields
\begin{equation}
\label{eq8}
f(z,m) \approx 
\frac{1}{Z_m} 
\sum_{\tau=0}^S \sum_{n=1}^R \delta_{mn} Y_n(\tau+1)
\exp\left\{ -\frac{[z-Y_n(\tau)]^2}{2h^2} \right\}  ,
\end{equation}
where the standard deviation $h$ is given by $h=s L^{-0.2}$, where
$s$ is the sample standard deviation of all $Y_n$ scores that 
belong to a particular month $m$ of the year 
(i.e., that show up on the sum and for which 
$\delta_{mn}=1$ holds --- 
these are the scores from which the density is estimated) 
and $L$ is the number of such 
scores~\cite{silverman86book,frank08aphysa}. 
 Moreover, 
$Z_m=
\sum_{\tau=0}^S \sum_{n=1}^R \delta_{mn} 
\exp\{ -[z-Y_n(\tau)]^2/(2h^2) \}$.
The interpolation modelling method (or regression model method) assumes that
$
f(z,m) \approx A_0(m)  + \sum_{j=1}^p A_j(m) z^j$.
For the relative small data sets that will be considered 
below, we will use the model that describes some dependency of $f$ on $z$ 
and features the smallest number of parameters. That is, we 
will consider the order $p=1$. In this case, equation~(\ref{eq2}) becomes
the linear regression model 
\begin{equation}
\label{eq10}
f(z,m) \approx A(m)  + B(m) z \ \Rightarrow \
Y_n(\tau+1)= A(m) + B(m) Y_n(\tau) + g(m) \, \epsilon(\tau)
\end{equation}
with $A=A_0$ and $B=A_1$. 
The intercept and slope parameters $A(m)$ and $B(m)$ 
can be estimated 
by fitting the linear regression model equation~(\ref{eq10}) 
to scatter plots of 
$Y_n(\tau+1)$ versus $Y_n(\tau)$ given for every month $m$.
In fact, the 
$Y_n(\tau+1)$ versus $Y_n(\tau)$ scatter plots are used
to determine $f$ for all three approximations defined by
equations~(\ref{eq6}), (\ref{eq8}) and (\ref{eq10}) since 
equations~(\ref{eq6}), (\ref{eq8}) and (\ref{eq10}) involve the data pairs
$Y_n(\tau+1)$ and $Y_n(\tau)$ 
for a fixed month $m$, that is, all pairs
$Y_n(\tau+1)$ and $Y_n(\tau)$ for which $n$ corresponds to a particular
month $m$ of the year.
Moreover, from equation~(\ref{eq5}) it follows that 
$g$ of the linear regression model equation~(\ref{eq10}) 
can be estimated from the root-mean-squared error RMSE of the
regression model like $g(m)=\mbox{RMSE}(m)/\sqrt{2}$.

\begin{figure}[!b]
\centerline{\includegraphics[width=0.75\textwidth]{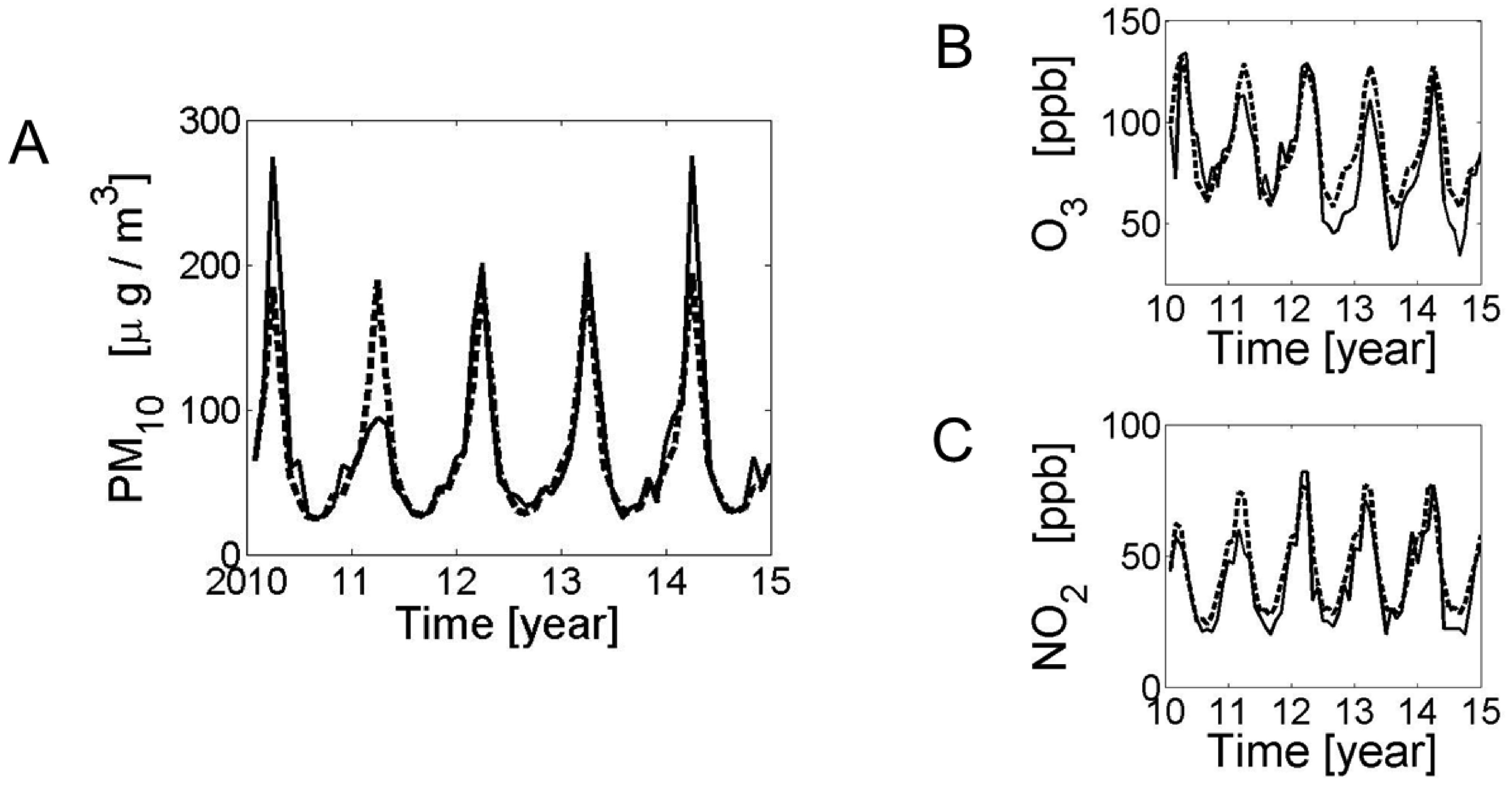}}
\caption{Extreme value pollutant concentrations (solid lines) 
measured in Chiangmai, North
  Thailand,  over the five years (i.e., 60 months) period from January 2010 to
  December 2014. Panels A, B, C show PM$_{10}$, O$_3$, and NO$_2$, 
respectively. Dashed lines show model fits obtained from the 
linear regression model
equation~(\ref{eq10}) in the deterministic case.}
\label{fig1}
\end{figure}

Data were taken from the Pollution Control Department (PCD) of
Thailand~\cite{PCD18}. Pollutant data 
for 
PM$_{10}$, 
O$_3$, and NO$_2$ in $N=60$
months from January 2010 to December 2014 were retrieved for the Provincial
Hall measurement station in Chiangmai. 
Figure~\ref{fig1} 
shows the pollutant time series. The station measured raw
PM$_{10}$ concentrations (in {\textmu}g/m$^3$) as averaged values for every day.
From the daily raw data, the PCD determined for each month 
the maximum scores. By contrast,    
O$_3$ and NO$_2$, 
raw concentrations (in ppb) were measured by the station every hour. 
From those hourly raw data, maximum scores of the day and 
maximum values for the month were determined. The monthly extreme value
data for 
PM$_{10}$, 
O$_3$, and NO$_2$ published on the PCD website~\cite{PCD18} 
were retrieved and analyzed. As mentioned above,
the study of extreme value data is of importance because extreme
pollutant concentrations are related to increased health 
risks~\cite{lid18,maji17,pope06}
and death 
rates~\cite{giri07,monterolorenzo11}.
All three pollutants 
PM$_{10}$, O$_3$, and NO$_2$
showed periodic annual patterns (i.e., seasonal effects), see 
figure~\ref{fig1}.
PM$_{10}$ extreme value concentrations 
peaked in the month of March. Similarly, O$_3$
extreme value 
concentrations reached maximum values during February, March, and April.
NO$_2$ extreme value concentrations were the largest in February and March.

For each pollutant trajectory $X(n)$, 
increment trajectories $Y_n(\tau)$ were derived 
for reference time points $n$
in the first three years (i.e.,  $n=1,\dots, R$ with $R=36$) such that each
trajectory covered a two years period (i.e., $\tau=0,\dots,S$ with $S=23$).
From the trajectories $Y_n$,
scatter plots for each month $m$ showing $Y_n(\tau+1)$ versus $Y_n(\tau)$ were
obtained. 
From the scatter plots, the drift functions $f$ were determined by means of
 the 3 
different
approximations defined by equations~(\ref{eq6}), (\ref{eq8}) and (\ref{eq10}).
Figure~\ref{fig2} shows the drift functions $f(Y,m)$ thus obtained for
PM$_{10}$ for the first four months of the year, January to April. 
The dashed lines
represent diagonals. For January and February, all three approximations of 
$f$ were above the diagonals 
indicating that PM$_{10}$ increment concentrations 
increased during those months. That is, if increments were positive in 
January (February),
then they tended to be positive and larger in magnitude in 
February (March). This describes the increase of the PM$_{10}$ pollutant
concentration $X(n)$ towards the peaks in March (see figure~\ref{fig1}A).
By contrast, for March and April, the drift functions
were found to be below the diagonals indicating the PM$_{10}$ increment
concentrations
decayed during those months. More precisely, if increments were positive in March (April), 
then they tended to be smaller (closer to zero) or negative in April (May). This corresponds to the decay
of the PM$_{10}$ pollutant concentration $X(n)$ from March to May 
[see figure~\ref{fig1}~(A) again].

\begin{figure}[!t]
\centerline{\includegraphics[width=0.69\textwidth]{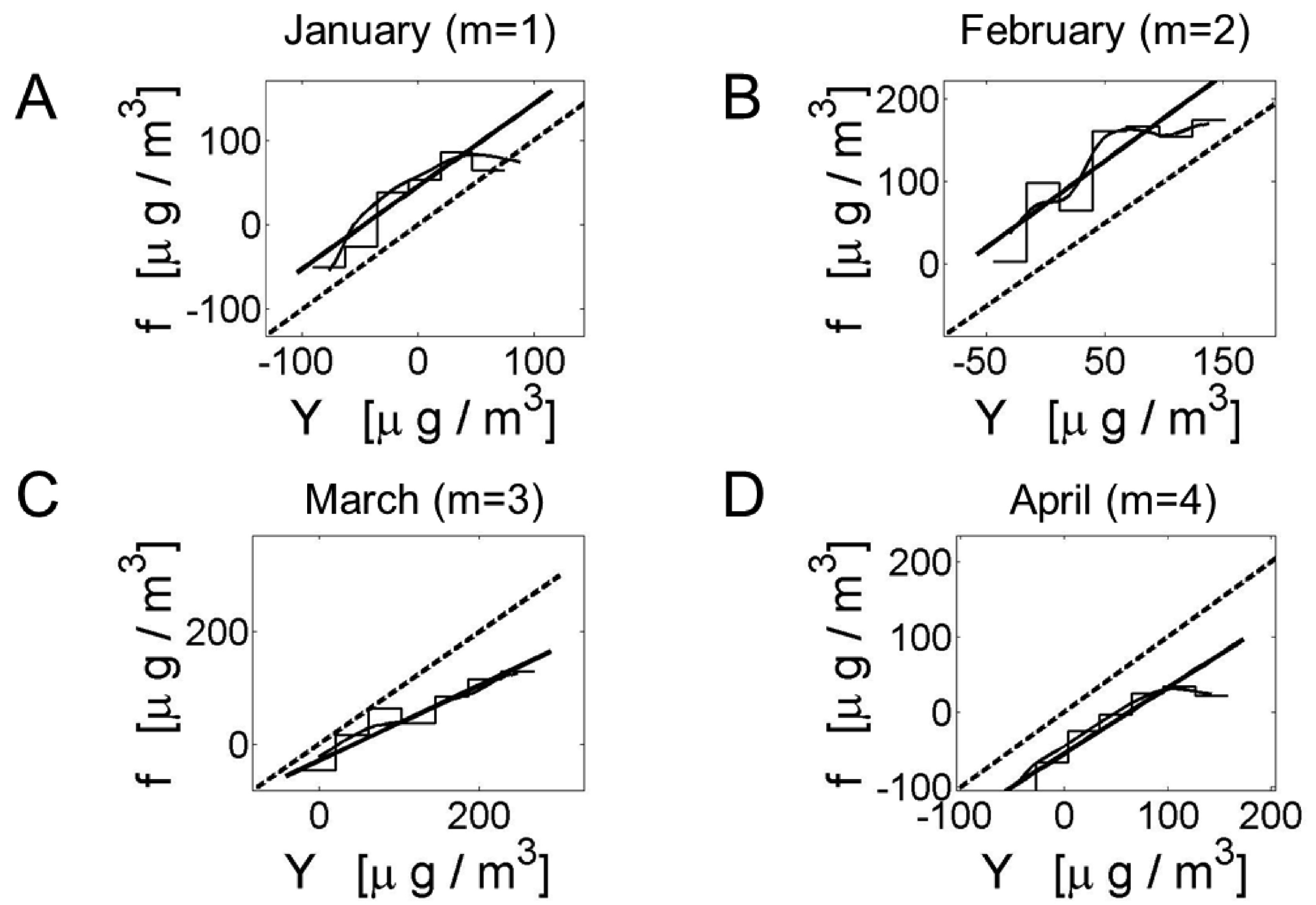}}
\caption{Drift functions $f(Y,m)$ for $Y=\rm{PM}_{10}$ 
extreme value concentrations
  determined for January, February, March, and April (panels A, B, C, and D)
  by means of three approximations: binning method
  (stair-step graphs), kernel density estimation method (solid smooth nonlinear
  lines),
  and
linear regression model (solid straight lines). Dashed lines represent
  diagonals.}
\label{fig2}
\end{figure}

By visual inspection of figure~\ref{fig2}, 
the kernel density estimation method has the advantage
to account for nonlinear characteristics of $f$ in a smooth fashion. It has
the disadvantage of being described by the whole data sets of 
$Y_n(\tau+1)$ and $Y_n(\tau)$ pairs 
that contribute to the relevant scatter
plots.  That is, each smooth function $f$ is characterized by a relatively
large set of parameters given in terms of 
$Y_n(\tau+1)$ and $Y_n(\tau)$ pairs.
The linear approximation has the advantage of being conveniently
characterized by two parameters $A$ and $B$. It has the disadvantage of not capturing nonlinear effects. The drift function approximation
by means of the binning method can be regarded as a compromise between the two
other approximations. The stair-case like approximate 
drift functions obtained from the
binning method account for nonlinearities. They 
can be described by a bin width
$\Delta y$, the bin centers $y_{k,\text{c}}=(y_{k+1}+y_{k})/2$, and the function
values $f_k$. Consequently, to describe $f$ with  $K$ bins, we need
$2K+1$
parameters. Therefore, the number of parameters to describe $f$ 
with the binning method is larger as
compared to the number of parameters that characterize 
the linear regression model but smaller as compared to the number of 
$Y_n(\tau+1)$ and $Y_n(\tau)$ parameters 
that are needed to approximate $f$ by 
means of the kernel density estimation method.

The linear regression model approximation, that has the advantage of requiring 
the smallest number of characterizing parameters, was determined in 
detail for all three pollutants. That is, the parameters $A$, $B$ and $g$
were estimated as described above. Figure~\ref{fig3} shows 
the model parameters thus obtained. 
For all three pollutants, the intercept parameter $A$ varied
across the months of the year in a characteristic pattern. 
For PM$_{10}$ and O$_3$ in January and February and 
for NO$_2$ in January it was positive and assumed the largest positive
values. 
Subsequently, in March and April (PM$_{10}$), April and May (O$_3$ and NO$_2$),
parameter $A$ was
negative
and assumed the largest-in-the-amount negative values. These patterns, 
as discussed
above in the context of PM$_{10}$ 
and figure~\ref{fig2}, describe the mechanism
that leads to the peaks in the original trajectories $X(n)$ around February,
March, and April, see figure~\ref{fig1}.
For the remaining months from May to December, the
intercept parameters $A$ 
were overall relatively small (i.e., close to zero).
PM$_{10}$ and O$_3$ showed exceptions from this rule in September, where the
parameter values
$A$
were positive and assumed 20\% (PM$_{10}$) and 45\% (O$_3$) of their
respective maximal positive $A$ parameters.
For all three pollutants, 
the slope parameter $B$ was found to
be relatively close to unity. 
For PM$_{10}$ and NO$_2$, the noise amplitude $g$ showed  clear 
seasonal peaks around
January, February, and March (PM$_{10}$) and March and April (NO$_2$).

\begin{figure}[!t]
\centerline{\includegraphics[width=0.78\textwidth]{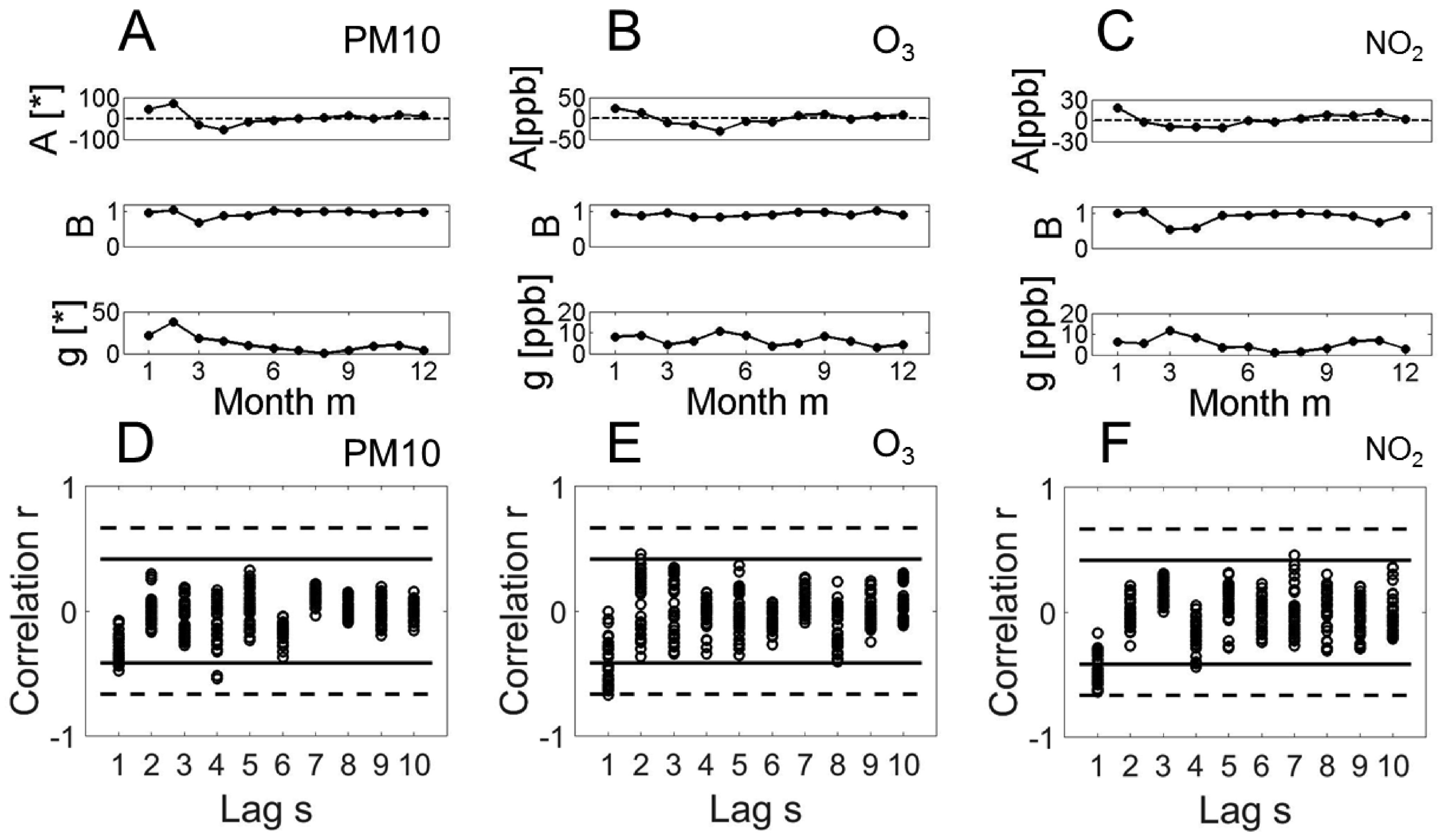}}  
\caption{Model parameters $A$, $B$, and $g$ as functions of month $m$ of the
  linear regression model equation~(\ref{eq10}) 
for the pollutants 
PM$_{10}$ (panel A),
  O$_3$ (panel B), and NO$_2$ (panel C). 
Here, $*$ in panel A means 
{\textmu}g/m$^3$. Panels D, E, F: Lag-$s$ 
autocorrelation coefficients of residuals for
PM$_{10}$ (panel D),
  O$_3$ (panel E), and NO$_2$ (panel F)  
with thresholds for statistical
significance (see text).}
\label{fig3}
\end{figure}

In order to validate the model, we tested the residuals $\epsilon$ occurring in 
equation~(\ref{eq10}).
We determined 
the first
ten lag-$s$ autocorrelation coefficients 
of the residuals  
for each trajectory
$Y_n(\tau)$.
The coefficients are shown in 
figure~\ref{fig3} (panels D, E, F) together with 
single-time-series thresholds \cite{diggle90book}
(solid lines) for statistically
significant 
and Bonferroni
adjusted~\cite{keppel04book} multiple-tests  thresholds
(dashed lines).
We found that some of the correlation coefficients (in particular, lag-1
correlation coefficients of O$_3$ and NO$_2$) 
violated the single-time-series criterion for being not
statistically significant. However, all correlation coefficients were found to
be within the boundaries of the multiple-tests thresholds. Residuals of
trajectories $Y_n(\tau)$ were also tested for violation of normality using
the Anderson-Darling normality test. For all PM$_{10}$ and
NO$_2$ trajectories $Y_n(\tau)$, the residuals did not violate the normality assumption. For O$_3$, 
the normality
assumption was violated in 4
out of $R=36$ trajectories $Y_n$. 
Overall, the correlation and normality tests 
supported the model 
assumptions.

We showed how to identify stochastic dynamical models for the evolution of air
pollutants on the basis of single, historical trajectories of pollutant
concentrations. To this end, we followed the earlier work on financial data and
considered pollutant increments rather than the raw pollutant data. 
In
addition, three different representation methods of the drift functions of
the dynamical models were used. In doing so, we derived three main results:
First, we found that all three representation methods were consistent with
each other, see figure \ref{fig2}. Second,
we were able to show that  experimentally observed 
annual air pollutant peaks were caused by
drift functions of physical-chemical air pollutant systems 
that change qualitatively from the pre-peak
months (e.g., January and February) to the post-peak months (e.g., March and
April), see figure~\ref{fig2} again. 
These qualitative changes in the drift functions
are assumed to reflect periodic changes in the physical-chemical laws
determining the evolution of the 
PM$_{10}$, O$_3$, and NO$_2$ pollutant concentrations.
Third, it was found that the linear approximation representation method of 
drift functions (which is the most parsimony method) is sufficient to reproduce
the emergence of the yearly pollutant peaks, see figure~\ref{fig1}.

\ukrainianpart
\title{Здійснення дрейф-дифузійного аналізу через дослідження турбулентних потоків та
	 динаміки частинок речовини смогу і забруднювачів повітря
	}
\author{T. Варапонгпісан\refaddr{label1}, Л. Інгрісвванг\refaddr{label1}, Т.Д. Френк\refaddr{label2,label3}}
\addresses{
	\addr{label1}
Факультет природничих наук, вiддiлення фiзики, унiверситет Касертсарт, Бангкок 10900, Таїланд
	\addr{label2} 
	CESPA, вiддiлення психологiї, Коннектикутський унiверситет, CT 06269, США
	\addr{label3} 
	Вiддiлення фiзики, Коннектикутський унiверситет, CT 06269, США
}

\newpage

\makeukrtitle

\begin{abstract}
Дрейф-дифузійний аналіз увійшов у фізику 
	як метод дослідження турбулентних
	потоків. 
	У даному дослідженні пропонується використовувати цей метод для ідентифікації
	базових динамічних моделей різних концентрацій твердих частинок смогу, озону і
	 діоксиду азоту. В роботі досліджуються дані з  Чіангмаї, найбільшого міста у північній частині  Таїланду, яке нещодавно стало свідком драматичних шпиталізацій, вочевидь
	пов'язаних з екстремальними рівнями забруднення повітря. Розглянуто
	три варіанти  дрейф-дифузійного аналізу
	(щільність ядра, бінінг та лінійне наближення).
	Показано, що всі три
	варіанти дають пояснення щорічним пікам забруднень впродовж першої половини року з урахуванням того, що параметри рівнянь 
	фізико-хімічної еволюції забруднювачів повітря періодично змінюються впродовж року.
	Отже, даний аналіз надає докази,
	що базові динамічні моделі
	трьох забруднювачів повітря, розглянутих у дослідженні, є явно залежними  від часу.

	\keywords дрейф-дифузійний аналіз, частинки речовини, забруднювачі повітря
	
\end{abstract}
\end{document}